\documentclass[a4paper,12pt]{article}
\usepackage{amssymb}
\usepackage{epsfig}
\usepackage{amsmath}

\author {Wen-Xiu Ma\thanks{Email: mawx@math.usf.edu, Fax: 1-813-9742700}
\\{ \small Department of Mathematics, University of South Florida, Tampa, FL 33620-5700, USA}
}
\title
{\sf Wronskians, Generalized Wronskians and Solutions to the
Korteweg-de Vries Equation} 
\date{\nonumber}
\begin{document}
\maketitle

\newcommand{\R}{\mathbb{R}}
\newcommand{\D}{\displaystyle}

\begin{abstract}
A bridge going from Wronskian solutions to generalized Wronskian
solutions of the Korteweg-de Vries equation is built. It is then
shown that generalized Wronskian solutions can be viewed as
Wronskian solutions.
% in the case of the Korteweg-de Vries equation.
The idea is used to generate positons, negatons and their
interaction solutions to the Korteweg-de Vries equation. Moreover,
general positons and negatons are constructed through the
Wronskian formulation. A few new exact solutions to the KdV
equation are explicitly presented as examples of Wronskian
solutions.

\end{abstract}

\numberwithin{equation}{section}

\newtheorem{thm}{Theorem}[section]
\newtheorem{Le}{Lemma}[section]

\setlength{\baselineskip}{18pt}
\def \part {\partial}
\def \be {\begin{equation}}
\def \ee {\end{equation}}
\def \bea {\begin{eqnarray}}
\def \eea {\end{eqnarray}}
\def \ba {\begin{array}}
\def \ea {\end{array}}
\def \si {\sigma}
\def \al {\alpha}
\def \la {\lambda}

\section{Introduction}

The Korteweg-de Vries (KdV) equation is one of the most important
models exhibiting the soliton phenomenon \cite{KruskalZ-PRL1965}.
Its multi-soliton solutions \cite{Hirota-PRL1971} can be expressed
by using a Wronskian determinant
\cite{Satsuma-JPSJ1979,FreemanN-PLA1983}. Matveev found that there
also exists another class of explicit solutions, called positons,
to the KdV equation, which can be presented by a generalized
Wronskian determinant \cite{Matveev-PLA1992}; and afterwards,
negatons were also constructed through taking advantage of the
generalized Wronskian determinant \cite{RasinariuSK-JPA1996}.

Solutions determined by the technique of Wronskian determinant and
the technique of generalized Wronskian determinant are called
Wronskian solutions and generalized Wronskian solutions,
respectively. Solitons are examples of Wronskian solutions, and
positons and negatons are examples of generalized Wronskian
solutions. A natural question to ask is whether there is any
interrelation between Wronskian solutions and generalized
Wronskian solutions. What kind of relation one can have if it
exists?

In this paper, we would like to build a bridge between Wronskian
solutions and generalized Wronskian solutions. This gives us a way
to obtain generalized Wronskian solutions through Wronskian
solutions to the KdV equation.
%Therefore, the answer is yes.
The basic idea is used to generate positons, negatons and their
interaction solutions to the KdV equation. Moreover, general
positons and negatons are constructed through the Wronskian
formulation. A few new exact solutions to the KdV equation are
explicitly presented as examples of Wronskian solutions.

\section{Bridge between Wronskians and generalized Wronskians}
\label{sec:generaltheory:kdv-gpn}

Let us specify the KdV equation as follows \be
u_{t}-6uu_x+u_{xxx}=0,\label{eq:kdv:kdv-gpn}\ee where (also in the
rest of the paper) $g_{y_1\cdots y_i}$ is the conventional
notation denoting the $i$-th order partial derivative $\partial
^ig/\partial y_1\cdots
\partial y_i$.
Hirota introduced
 the transformation \cite{Hirota-PRL1971}:
 \be u=-2\partial _x^2 \ln f= -\frac
{2(ff_{xx}-f_x^2)}{f^2},
\label{eq:transformationbetweenKdVandbilinearKdV:kdv-gpn}\ee
%which can be expressed as $-2(\textrm{ln}f)_{xx}$ as a convenient notation,
between the KdV equation (\ref{eq:kdv:kdv-gpn}) and the following
bilinear equation
 \be (D_xD_t+D_x^4)f\cdot
f=f_{xt}f-f_tf_x+f_{xxxx}f-4f_{xxx}f_x+3f_{xx}^2=0,
\label{eq:bkdv:kdv-gpn}\ee where $D_x$ and $D_t$, called the
Hirota operators, are defined by \be
f(x+k,t+h)g(x-k,t-h)=\sum_{i,j=0}^\infty \frac 1 {i!j!}(D_x^iD_t^j
f\cdot g)k^ih^j.\ee

Solutions to the above bilinear KdV equation
(\ref{eq:bkdv:kdv-gpn}) can be given by the Wronskian determinant
\cite{Satsuma-JPSJ1979,FreemanN-PLA1983}: \be W(\phi
_1,\phi_2,\cdots,\phi_{N})=\left|
\ba {cccc}\phi _1^{(0)}&\phi _1^{(1)}&\cdots &\phi_1^{(N-1)}\\
\phi _2^{(0)}&\phi_2^{(1)}&\cdots
 &\phi_2^{(N-1)}\\ \vdots &\vdots & \ddots & \vdots \\
\phi _{N}^{(0)}&\phi_{N}^{(1)}&\cdots
 &\phi_N^{(N-1)}\ea
\right |,\ N\ge 1,\label{eq:wronskian:complexitonkdv}\ee where \be
\phi_i^{(0)}=\phi_i,\, \phi _i^{(j)}=\frac {\part ^j}{\part
x^j}\phi_i,\ j\ge 1,\,1\le i\le N.\ee
 Sirianunpiboon et al.
\cite{SirianunpiboonHR-PLA1988} furnished the following conditions
\be -\phi_{i,xx}=\sum_{j=1}^i\lambda _{ij} \phi_j,\
\phi_{i,t}=-4\phi_{i,xxx},\ 1\le i\le N,
\label{eq:lowertriangularcaseofsufficentconditions:kdv-gpn} \ee
where $\lambda_{ij}$ are arbitrary real constants, to make the
Wronskian determinant a solution to the bilinear KdV equation
(\ref{eq:bkdv:kdv-gpn}), and showed that rational function
solutions and their interaction solutions with multi-soliton ones
to the KdV equation can be obtained this way.

Matveev found \cite{Matveev-PLA1992} that there exists so-called
generalized Wronskian solutions to the bilinear KdV equation
(\ref{eq:bkdv:kdv-gpn}), which read as \be W(\phi,\partial _k
\phi,\cdots,
\partial _k ^{n}\phi)
, \ \partial _k^i=\frac {\partial ^i}{\partial k^i } ,\ 1\le i\le
n,\label{eq:gwd:kdv-gpn}\ee where the function $\phi$ satisfies
\[-\phi_{xx}=k^2 \phi, \
\phi_{t}=-4\phi_{xxx},\ k\in \R . \] The resulting solutions to
the KdV equation (\ref{eq:kdv:kdv-gpn}) are called positons, since
the corresponding eigenfunctions are associated with positive
eigenvalues of the Schr\"odinger spectral problem.

Generally, the generalized Wronskian determinant
(\ref{eq:gwd:kdv-gpn}) gives rise to a solution to the bilinear
KdV equation (\ref{eq:bkdv:kdv-gpn}), provided that the function
$\phi$ satisfies
 \be -\phi_{xx}=\alpha(k) \phi, \
\phi_{t}=-4\phi_{xxx},\label{eq:scforWDofKdV:kdv-gpn}
 \ee
with $\alpha$ being an arbitrary function of $k\in\R $. Observe
that if we have (\ref{eq:scforWDofKdV:kdv-gpn}),
 then the function $\phi$ also satisfies
 \be  -\bigl({\partial_
k^m\phi }\bigr)_{xx}=\sum_{i=0}^m {m\choose i}({\partial
_k^i\alpha})( {\partial _k^{m-i}\phi }) ,\
  \bigl(
{\partial_k^m \phi}\bigr)_{t} =-4 \bigl(
{\partial_k^m\phi}\bigr)_{xxx},\ m\ge 0.\ee
 Upon setting \be \psi _{i+1}=
\frac 1{i!}\frac {\partial  ^i\phi}{\partial k^i},\
\alpha_{i+1}=\frac 1{i!}\frac {\partial  ^i\alpha}{\partial k^i},\
i\ge 0,\label{eq:tsofphiandpsi:kdv-gpn}\ee it follows that the
functions $\psi_i$, $1\le i\le n+1$, satisfy a lower triangular
system of second-order differential equations: \be \left\{\ba {l}
-\psi_{1,xx}=\alpha_1\psi_1 ,\vspace{2mm}\\
 -\psi_{2,xx}=\alpha_2 \psi_1
+\alpha_1\psi_2,\vspace{2mm}\\
\quad \cdots\cdots   \vspace{2mm}\\
 -\psi_{n+1,xx}= \alpha_{n+1}\psi_{1}+\alpha_{n}\psi_{2}+\cdots+
\alpha_1\psi_{n+1}.
 \ea \right.\label{eq:alowertriangularsystem:kdv-gpn}\ee
 This is to say,
 \be
- \Psi_{xx}=\Lambda \Psi, \
 \Lambda =
 \left[\ba {cccc} \alpha_1&&&0 \vspace{2mm}\\
\alpha_2&\alpha_1& & \vspace{2mm}\\
\vdots & \ddots&\ddots &  \vspace{2mm}\\
\alpha_{n+1}&\cdots &\alpha_2 &\alpha_1
  \ea \right],\
  \Psi=\left[
  \ba {c} \psi_1\vspace{2mm}\\ \psi_2,\vspace{2mm}\\
  \vdots \vspace{2mm}\\
  \psi_{n+1} \ea
  \right] ,
  \ee
  which is a special case of the conditions (\ref{eq:lowertriangularcaseofsufficentconditions:kdv-gpn}).
  Obviously, under the transformation (\ref{eq:tsofphiandpsi:kdv-gpn}), the above system of differential equations,
  together with
\[\psi_{i,t}=-4\psi_{i,xxx},\ 1\le i\le n+1,\] is equivalent to the conditions
  (\ref{eq:scforWDofKdV:kdv-gpn}).
Therefore, summing up, there exists a bridge going from the
Wronskian to the generalized Wronskian:
 \be W(\psi_1,\psi_2,\cdots,\psi_{n+1})=(\prod_{i=1}^n \frac 1 {i!}) W(\phi,
\partial _k \phi,\cdots ,\partial_k ^n\phi).\label{eq:bridge:kdv-gpn}\ee
The constant factor in (\ref{eq:bridge:kdv-gpn}) does not affect
the final solution determined by
(\ref{eq:transformationbetweenKdVandbilinearKdV:kdv-gpn}), i.e.,
we have \be  u=-2\partial_x^2\ln
W(\psi_1,\psi_2,\cdots,\psi_{n+1})= -2\partial_x^2\ln W(\phi,
\partial _k \phi,\cdots ,\partial_k ^n\phi).\label{eq:relationbetweenWSandGWS:Kdv-gpn} \ee
This implies that \be u=-2\partial_x^2\ln W(\phi,
\partial _k \phi,\cdots ,\partial_k ^n\phi) \ee
gives a solution to the KdV equation (\ref{eq:kdv:kdv-gpn}) if
(\ref{eq:scforWDofKdV:kdv-gpn}) holds, and
%Now it follows from (\ref{eq:relationbetweenWSandGWS:Kdv-gpn})
 that all such
generalized Wronskian solutions to the KdV equation can be
obtained through the Wronskian formulation. However,
(\ref{eq:alowertriangularsystem:kdv-gpn}) can have other solutions
besides $(\phi,
\partial _k \phi,\cdots ,\partial_k ^n\phi)$ with $\phi$ solving
(\ref{eq:scforWDofKdV:kdv-gpn}), and thus not all Wronskian
solutions are of generalized Wronskian type.

\section{Positons, negatons and their interaction solutions}

Two particular classes of generalized Wronskian solutions to the
KdV equation are positons and negatons. It is known  that positons
of order $n$ are represented by using the generalized Wronskian
determinant \cite{Matveev-PLA1992}: \be f=W(\phi,\partial _k
\phi,\cdots,
\partial _k^n\phi ),\ \phi=\cos (kx+4k^3t+\gamma(k) ), \label{eq:eigenfunctionsforpositon1:kdv-gpn}\ee or \be
f=W(\phi,\partial _k \phi,\cdots,
\partial _k^n\phi ),\ \phi=\sin (kx+4k^3t+\gamma(k) ),\label{eq:eigenfunctionsforpositon2:kdv-gpn} \ee  where
$\gamma $ is an arbitrary function of $k$; and that
 negatons of order $n$, by using the generalized Wronskian determinant
\cite{RasinariuSK-JPA1996}: \be f= W(\phi,\partial _k \phi,\cdots,
\partial _k^n\phi ),\ \phi=\textrm{cosh}({kx-4k^3t+\gamma(k) }),\label{eq:eigenfunctionsfornegaton1:kdv-gpn}
\ee or \be f=W(\phi,\partial _k \phi,\cdots, \partial _k^n\phi ),\
\phi =\textrm{sinh}({kx-4k^3t+\gamma(k) }),
\label{eq:eigenfunctionsfornegaton2:kdv-gpn}\ee  where $\gamma $
is an arbitrary function of $k$ as well. Note that two kinds of
positons are equivalent to each other, due to the existence of an
arbitrary function $\gamma(k)$. But two kinds of negatons are
functionally independent.

 In the case of positons,
we have
 \be \alpha(k)=k^2, \ k\in
\R \ee in the conditions
  (\ref{eq:scforWDofKdV:kdv-gpn}),
 which implies that the Schr\"odinger
 %spectral problem
 operator $-\partial ^2/\partial x^2+u$
 with zero potential
  has a positive
 eigenvalue.
Further, we have \be \alpha_1=k^2,\ \alpha_2=2k,\ \alpha_3=1,\
\alpha_i=0,\ i\ge 4,\ee
 and the corresponding coefficient matrix becomes
\be \Lambda = \left[\ba {ccccc}
k^2&&&&0 \vspace{2mm}\\
2k&k^2& & &\vspace{2mm}\\
1& 2k &k^2& &\vspace{2mm}\\
& \ddots & \ddots & \ddots &\vspace{2mm}\\
0 & &1 &2k &k^2 \ea \right].\ee
 In the case of negatons, we
have
 \be
 \alpha(k)=-k^2, \ k\in
\R \ee in the conditions (\ref{eq:scforWDofKdV:kdv-gpn}), which
implies that the Schr\"odinger operator $-\partial ^2/\partial
x^2+u$ with zero potential has a negative
 eigenvalue.
Thus, we have \be \alpha_1=-k^2,\ \alpha_2=-2k,\ \alpha_3=-1,\
\alpha_i=0,\ i\ge 4,\ee
   and the corresponding coefficient matrix reads as
\be \Lambda = \left[\ba {ccccc}
-k^2&&&&0 \vspace{2mm}\\
-2k&-k^2& & &\vspace{2mm}\\
-1&-2k &-k^2& &\vspace{2mm}\\
 & \ddots & \ddots & \ddots &\vspace{2mm}\\
0 & &-1 &-2k &-k^2 \ea \right].\ee Therefore, all positons and
negatons can be presented through the Wronskian formulation.

The manipulation in the previous section also allows us to
generate interaction solutions among positons and negatons. Let us
choose the functions $\phi_i$, $1\le i\le m$, among the functions
\be \cos (k_ix+4k_i^3t+\gamma_i(k_i) ) ,\ \sin
(k_ix+4k_i^3t+\gamma_i(k_i)
),\label{eq:positoneigenfunctions:kdv-gpn}\ee and the functions
\be \textrm{cosh}({k_ix-4k_i^3t+\gamma_i(k_i) }),\
\textrm{sinh}({k_ix-4k_i^3t+\gamma_i(k_i) }),
\label{eq:negatoneigenfunctions:kdv-gpn} \ee where the $k_i$'s are
arbitrary constants and the $\gamma_i$'s are arbitrary
  functions.
 Then we have a
new class of exact solutions to the KdV equation
(\ref{eq:kdv:kdv-gpn}): \be u_m=-2\partial _x^2\ln f =-\frac
{2(ff_{xx}-f_x^2)}{f^2},
 \ee
where the function $f$ is given by  \be f=W(\phi_1,\partial
_k\phi_1,\cdots,
\partial _k^{n_1}\phi _1;\cdots;
\phi_m,\partial _k\phi_m,\cdots,
\partial _k^{n_m}\phi_m
)\label{eq:isofpn:kdv-gpn}\ee with arbitrary non-negative integers
$n_1,n_2,\cdots,n_m$. This is an abroad class of interaction
solutions among positons and negatons to the KdV equation
(\ref{eq:kdv:kdv-gpn}).

In particular, let us fix \be \phi_1= \sin (kx+4k^3t+\gamma),\
\phi_2= \sinh (kx-4k^3t+\gamma), \ee  where $k$ and $\gamma  $ are
two arbitrary constants. Then the Wronskian determinant $
f=W(\phi_1,\phi_2)$ becomes
 \be f=W(\phi_1,\phi_2)=
k(\sin \xi_+ \cosh \xi_--\sinh \xi_- \cos \xi_+),
 \ee and the corresponding interaction solution between a
 simple positon and a simple negaton reads as
 \begin{eqnarray}
u&=&-2\partial _x^2\ln f =-2\partial _x^2\ln W(\phi_1,\phi_2) \nonumber \\
&=& \frac { 4 k^2 (\sinh ^2 \xi_- - \sin ^2\xi_+ ) }{(\sin \xi_+
\cosh \xi_- -\sinh \xi_- \cos \xi_+ )^2}, \label{eq:pn:kdv-gpn}
\end{eqnarray}
 where \be \xi _+=kx+4k^3t+\gamma,\ \xi_-= kx-4k^3t+\gamma. \ee
The graphs of the solution in two cases
 are depicted with grid$=[60,60]$ in Figure \ref{fig:pn:kdv-gpn}, which show the distribution of some
 singularities.
 %\begin{figure}[h] {\epsfig{figure=kdv-pn1.ps, width=7cm, height=5cm}}
 %{\epsfig{figure=kdv-pn2.ps, width=7cm, height=5cm}}
% \caption{Interaction solution - $k=1,\,\gamma =1$
%(left) and $k=2,\,\gamma=-2$ (right)} \label{fig:pn:kdv-gpn}
%\end{figure}

\section{General positons and negatons}

In this section, we are going to present two classes of general
positons and negatons to the KdV equation (\ref{eq:kdv:kdv-gpn}),
which provide us with new exact solutions to the KdV equation
(\ref{eq:kdv:kdv-gpn}).

Let us first start from
 \be -\phi_{+,xx}=k^2 \phi_+, \
\phi_{+,t}=-4\phi_{+,xxx},\ k\in \R
.\label{sec:gpconditions:kdv-gpn}
 \ee
 A general solution to
 the system (\ref{sec:gpconditions:kdv-gpn}) is given by
 \be \phi_+(k)
=c(k)
 \cos(kx+4k^3t)+ d(k)\sin (kx+4k^3t)
  , \label{eq:phi(k)forgeneralpositon:kdv-gpn} \ee
 where $c$ and $d$ are two arbitrary functions of $k$.
Then based on our construction in Section
\ref{sec:generaltheory:kdv-gpn}, we obtain a class of exact
solutions to the KdV equation (\ref{eq:kdv:kdv-gpn}): \be
u=-2\partial _x^2 \ln W(\phi_+(k),\partial _k\phi_+(k),\cdots,
\partial _k^n\phi_+(k) ),
 \ee
 where $\phi_+(k)$ is given by
 (\ref{eq:phi(k)forgeneralpositon:kdv-gpn}).
Such solutions correspond to the positive eigenvalue of the
Schr\"odinger spectral problem, and simple positons determined by
(\ref{eq:eigenfunctionsforpositon1:kdv-gpn}) and
(\ref{eq:eigenfunctionsforpositon2:kdv-gpn}) are just their two
examples in the cases of \[ c=\cos(\gamma (k)),\ d=-\sin (\gamma
(k)) \quad \textrm{and}\quad c=\sin(\gamma (k)),\ d=\cos (\gamma
(k)) .\] If we choose $c$ and $d$ to be constants, then the
Wronskian determinant involving the first-order derivative
$\partial _k\phi_+$
becomes \begin{eqnarray} &&W(\phi_+(k),\partial _k\phi_+(k)) \nonumber \\
&=&
 -\frac 12 (c^2-d^2)\sin (2\xi_+)+cd\cos
(2\xi_+)-(c^2+d^2)kx-12(c^2+d^2)k^3t , \quad \end{eqnarray}
 and the corresponding general positon
of order 1 reads as
\begin{eqnarray}
 u&=&-2\partial _x^2 \ln W(\phi_+(k),\partial _k\phi_+(k)) \nonumber \\
 &=& \D \frac {
4k^2(c^2+d^2)[(c^2-d^2)xk+12(c^2-d^2)k^3t+2cd]\sin
(2\xi_+)}{[-\frac 12 (c^2-d^2)\sin (2\xi_+)+cd\cos
(2\xi_+)-(c^2+d^2)kx-12(c^2+d^2)k^3t]^2}\nonumber \\
&&-\D \frac {
 4k^2(c^2+d^2)[(2cdkx+24cdk^3t-c^2+d^2) \cos
(2\xi_+)-(c^2+d^2)] }{[-\frac 12 (c^2-d^2)\sin (2\xi_+)+cd\cos
(2\xi_+)-(c^2+d^2)kx-12(c^2+d^2)k^3t]^2}
 ,\qquad \  \  \label{eq:sgp:kdv-gpn}\end{eqnarray}
where $c$, $d$ and $k$ are arbitrary constants and $\xi_+$ is
given by \be \xi_+= k x+4k^3t. \ee  The graphs of the solution in
two cases
 are depicted with grid$=[60,60]$ in Figure \ref{fig:gp:kdv-gpn}, which exhibit some
 singularities of the solution.
 %\begin{figure}[h] {\epsfig{figure=kdv-gp1.ps, width=7cm, height=5cm}}
 %{\epsfig{figure=kdv-gp2.ps, width=7cm, height=5cm}}
 %\caption{General positon - $k=1,\,c=-d=1$
%(left) and $k=3,\,c=d=1$ (right)} \label{fig:gp:kdv-gpn} \end{figure}

Let us second start from
 \be -\phi _{-,xx}=-k^2 \phi _-, \
\phi _{-,t}=-4\phi _{-,xxx},\ k\in \R
.\label{sec:gnconditions:kdv-gpn}
 \ee
 A general solution to
 the system (\ref{sec:gnconditions:kdv-gpn}) is given by
 \be \phi _-(k)
=c(k)
 \textrm{exp}(kx-4k^3t)+ d(k)\textrm{exp}(-kx+4k^3t)
  , \label{eq:psi(k)forgeneralnegaton:kdv-gpn} \ee
 where $c$ and $d$ are two arbitrary functions of $k$.
Then similarly, based on our construction in Section
\ref{sec:generaltheory:kdv-gpn}, we obtain another class of exact
solutions to the KdV equation (\ref{eq:kdv:kdv-gpn}): \be
u=-2\partial _x^2 \ln W(\phi _-(k),\partial _k\phi _-(k),\cdots,
\partial _k^n\phi _-(k) ),
 \ee
 where $\phi _-(k)$ is given by
 (\ref{eq:psi(k)forgeneralnegaton:kdv-gpn}).
Such solutions correspond to the negative eigenvalue of the
Schr\"odinger spectral problem, and simple negatons determined by
(\ref{eq:eigenfunctionsfornegaton1:kdv-gpn}) and
(\ref{eq:eigenfunctionsfornegaton2:kdv-gpn}) are just their two
special examples in the cases of
\[ c=\frac 12 \textrm{e}^{\gamma(k)},\ d=\frac 12 \textrm{e}^{-\gamma(k)} \quad \textrm{and}\quad
c=\frac 12 \textrm{e}^{\gamma(k)},\ d=-\frac 12
\textrm{e}^{-\gamma(k)} .\]
 If
we choose $c$ and $d$ to be constants, then the Wronskian
determinant involving the first-order derivative $\partial
_k\phi_-$
becomes
\begin{equation}  W(\phi _-(k),\partial _k\phi _-(k))=
c^2 \textrm{e}^{2 \xi_- }-d^2 \textrm{e}^{-2 \xi_-}+4cdk( x-12 k^2
t) ,
\end{equation}
 and the
corresponding general negaton of order 1 reads as
\begin{eqnarray}
 u&= &-2\partial _x^2\ln W(\phi _-(k),\partial _k\phi
 _-(k))\nonumber \\
 &=&
 \frac {-32 c d  k^2 [c^2( kx-  12  k^3 t -1)\textrm{e}^{2
\xi_- } -d^2(kx-12k^3t+1) \textrm{e}^{-2\xi_-} -2 c d]}{[c^2
\textrm{e}^{2 \xi_- }-d^2 \textrm{e}^{-2 \xi_-}+4cdk( x-12 k^2 t)
]^2}
 ,\qquad \label{eq:sgn:kdv-gpn}
\end{eqnarray}
 where $c$, $d$
and $k$ are arbitrary constants and $\xi_-$ is given by \be
\xi_-=kx-4k^3t. \ee  The graphs of the solution in two cases
 are depicted with grid$=[60,60]$ in Figure \ref{fig:gn:kdv-gpn}, which show the distribution of some
 singularities of the solution.
 %\begin{figure}[h] {\epsfig{figure=kdv-gn1.ps, width=7cm, height=5cm}}
 %{\epsfig{figure=kdv-gn2.ps, width=7cm, height=5cm}}
 %\caption{General negaton - $k=1,\,2c=d=2$
%(left) and $k=2,\,c=-d=1$ (right)} \label{fig:gn:kdv-gpn}
%\end{figure}

\section{Conclusion and remarks}

On one hand, a bridge between Wronskian solutions and generalized
Wronskian solutions to the KdV equation was built. It gives us a
way to obtain generalized Wronskian solutions simply from
Wronskian determinants. The basic idea was used to generate
positons, negatons and their interaction solutions to the KdV
equation through the Wronskian formulation. A specific interaction
solution between two simple positon and negaton to the KdV
equation (\ref{eq:kdv:kdv-gpn}) was given by
(\ref{eq:pn:kdv-gpn}). On the other hand, general positons and
negatons were also presented through the Wronskian formulation.
They provide new examples of Wronskian solutions. Two new specific
solutions of general positons and negatons to the KdV equation
(\ref{eq:kdv:kdv-gpn}) were given by (\ref{eq:sgp:kdv-gpn}) and
(\ref{eq:sgn:kdv-gpn}).

There are also interaction solutions between positons and solitons
\cite{Matveev-PLA1992b}.
 Such solutions
can also be constructed through the Wronskian determinant
 \be
%u=-\frac {2(ff_{xx}-f_x^2)}{f^2}, \
f=W(\phi,\partial_k\phi,\cdots,\partial _k^n; \phi_1,\cdots,
\phi_N), \label{eq:isofps:kdv-gpn}\ee
 where
the function $\phi$ is chosen from the functions in
(\ref{eq:positoneigenfunctions:kdv-gpn}) and the functions
$\phi_i$ are chosen as \[\ba {l}
\phi_i=\textrm{cosh}({k_ix-4k_i^3t}+\gamma_i) ,\
\gamma_i=\textrm{const.}, \ \textrm{if}\   i \
\textrm{odd},\vspace{2mm}\\
\phi_i=\textrm{sinh}({k_ix-4k_i^3t}+\gamma_i),\
\gamma_i=\textrm{const.}, \ \textrm{if}\  i\ \textrm{even}.\ea
\] Moreover, if we choose the function $\phi$ from the functions in
(\ref{eq:negatoneigenfunctions:kdv-gpn}), then the Wronskian
determinant given by (\ref{eq:isofps:kdv-gpn}) generates
 interaction
solutions between negatons and solitons to the KdV equation
(\ref{eq:kdv:kdv-gpn}). Combining the constructions in
(\ref{eq:isofpn:kdv-gpn}) and (\ref{eq:isofps:kdv-gpn}) will give
rise to more general interaction solutions among positons,
negatons and solitons. We believe that such an idea of
constructing interaction solutions should work for other soliton
equations, especially for the perturbation KdV equations
\cite{MaF-PLA1996}.

Finally, we point out that there is also another class of explicit
exact solutions to the KdV equation (\ref{eq:kdv:kdv-gpn}), called
complexitons \cite{Ma-PLA2002}. One-complexiton is given by \bea
u&=&
%-2\partial _x^2 \ln W(\phi_{11},\phi_{12}) \nonumber\\ &=&
 \D \frac {-4\beta ^2\bigl[1+\cos(2\delta(x-\bar \beta
t)+\kappa)\cosh (2\Delta(x+\bar \alpha t)+\gamma)
\bigr]}{\bigl[\Delta  \sin (2\delta(x-\bar \beta t)+\kappa)+\delta
\sinh (2\Delta(x+\bar \alpha
t)+\gamma)\bigr ]^2} \nonumber \\
&& + \D \frac {4 \alpha \beta \sin (2\delta(x-\bar \beta
t)+\kappa)\sinh (2\Delta(x+\bar \alpha t)+\gamma) }{\bigl[\Delta
 \sin (2\delta(x-\bar \beta t)+\kappa)+\delta  \sinh
(2\Delta(x+\bar \alpha t)+\gamma)\bigr ]^2},
\label{eq:onecomplexitonofKdV:ma144} \eea where $\alpha ,\,\beta
>0,\, \kappa$ and
$\gamma$ are arbitrary real constants, and $\Delta ,\, \delta ,\,
\bar {\alpha },$ and $\bar {\beta }$ are given by
 \bea && \Delta =\sqrt{\frac {\sqrt{\alpha ^2+\beta^2}-\alpha }{2}},\ \delta =\sqrt{\frac {\sqrt{\alpha
^2+\beta ^2}+\alpha }{2}}, \nonumber  \\ &&  \bar {\alpha
}=4\sqrt{\alpha ^2+\beta^2}+8\alpha  ,\ \bar {\beta
}=4\sqrt{\alpha^2+\beta^2}-8\alpha . \nonumber \eea It requires a
generalization of the conditions
(\ref{eq:lowertriangularcaseofsufficentconditions:kdv-gpn}) to
construct interaction solutions among rational solutions,
solitons, positons, negatons and complexitons, which will be
discussed elsewhere. Positons, most of negatons (except solitons)
and complexitons exhibit different singularities. A general theory
on singularity of the KdV equation needs to be explored.

\noindent {\bf Acknowledgments:} The author greatly appreciates
stimulating discussions with R. Conte, C. Gilson, Y. S. Li, K.
Maruno and M. Pavlov. The work was in part supported by grants
from the Research Grants Council of Hong Kong.

\noindent {\bf Note:} The complete Latex file and used figure ps
files can be found at

\centerline{http://www.math.usf.edu/$\sim$mawx/kdv-gpn.htm.}

\end{document}